# Smartphone scene generator for efficient characterization of visible imaging detectors

Michael Bottom,[a] Leo S. Neat,[b] Leon K. Harding,[a] Patrick Morrissey,[a] Seth R. Meeker,[a] Richard T. Demers[a]

[a]Jet Propulsion Laboratory, California Institute of Technology, Pasadena, CA, 91109 USA

[b]University of California, Santa Cruz, 1156 High St, Santa Cruz, CA 95064

## ABSTRACT

Full characterization of imaging detectors involves subjecting them to spatially and temporally varying illumination patterns over a large dynamic range. Here we present a scene generator that fulfills many of these functions. Based on a modern smartphone, it has a number of good features, including the ability to generate nearly arbitrary optical scenes, high spatial resolution (13 um), high dynamic range ($\sim 10^4$), near-Poisson limited illumination stability over time periods from 100 ms to many days, and no background noise. The system does not require any moving parts and may be constructed at modest cost. We present the optical, mechanical, and software design, test data validating the performance, and application examples.

## 1. INTRODUCTION

Large-format, high sensitivity visible imaging detectors are now becoming common in many fields of science. For the most stringent applications such as medical imaging and astronomy, careful characterization of these devices is required to control systematic errors and noise. Issues like dark current, readout noise, clock-induced charge, charge transfer efficiency, pixel quantum efficiency non-uniformities, and spatially dependent losses caused by charge traps are important to understand in order to determine the noise that is added by the detector.

Modeling noise and systematics can be a useful first step, but models of detectors may not contain all the relevant physics to characterize systematic errors. Measuring these systematics is time consuming and challenging. Often, extensive optical test devices are used,[1,2] and as these detectors grow larger, the cost and complexity of the test hardware grows and independent assemblies may be required. For example, the optics typically used for measuring the flat field response, such as an integrating sphere, are quite different from those required to measure spatially dependent point-spread-function distortions, such as a pinhole grid. Having independent assemblies or configurations increases the chances of systematic errors.

In this paper we present a system developed to characterize electron-multiplying CCD (EMCCD) detectors for the Wide-Field Infrared Survey Telescope (WFIRST) coronagraph,* an instrument designed to image planets orbiting nearby stars.[3] The demands placed on such exoplanet imaging detectors are extreme. Planets are fainter than their stars by factors of millions to billions, and even after starlight suppression through a coronagraph and wavefront control system, dynamic ranges of thousands or more will be present on the detector. Planets are also quite faint in absolute terms, with fluence rates of tens of photons per square meter per hour or less being common, so photon-counting abilities are needed, as well as very low dark current, readout noise, and clock-induced charge. However, this calibration system can be used with a wide range of detectors. It has a number of useful features including:

- The ability to generate nearly arbitrary optical scenes through software alone (no hardware changes), including flatfields, pinhole grids, and other test patterns

---

Send correspondence to mbottom@jpl.nasa.gov

*The decision to implement WFIRST will not be finalized until NASA's completion of the National Environmental Policy Act (NEPA) process. This document is being made available for information purposes only.

- Very high spatial resolution (20 $\mu$m), with the ability to generate ~2 million independent PSFs in a 1cm$^2$ imaging area

- High single-frame dynamic range (~$10^4$)

- The ability to generate temporally changing illumination patterns, such as video scenes

- Near Poisson-limited stability in illumination with no stray or background light

- Diffraction-limited performance over the field of view

- Modest cost and complexity (~$2000; off-the-shelf parts)

- Simple setup, alignment, and tolerances

## 2. OPTOMECHANICAL DESIGN

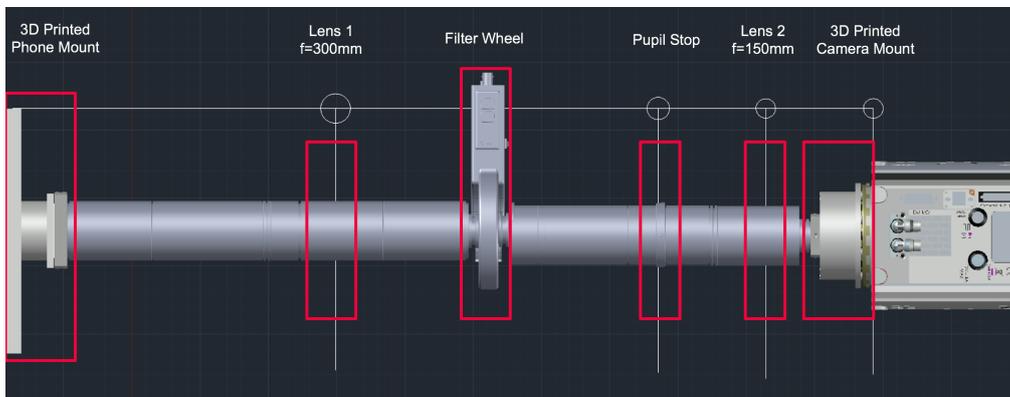

Figure 1: 2d layout of the scene generator with a test camera (Andor iXon), showing various components.

The scene generator is a simple two-lens optical system that re-images a smartphone screen onto the detector. A drawing of the scene generator interfaced to a test camera is shown in Figure 1.

### 2.1 Smartphone OLED screens

The key enabling technology for the scene generator is the introduction of high pixel density organic light-emitting diode (OLED) displays on smartphones.[4] Previous models of smartphones used conventional liquid crystal displays (LCDs), in which each pixel is coated with a particular red, green, or blue filter in a Bayer-like pattern. These pixels are back-illuminated by a white light, which can be reduced in intensity but not turned off. This renders conventional LCDs unacceptable, as even low levels of ambient background are easily visible with modern optical detectors. Newer smartphones models use OLEDs, which have a different structure. In these pixels, an electroluminescent layer of an organic compound emits light in response to applied current. Each OLED emits an intrinsic spectrum without need for a filter or backlight (see Figure 2). With OLEDs, individual pixels will emit no light when the applied current drops below a threshold value. Note that the phones have such high pixel density that the resolution of the human eye cannot distinguish the individual RGB elements of each pixel without the aid of magnifying optics.

OLED smartphone screens make useful media for projecting scenes. The current screen pixels sizes of 12-20 $\mu$m and pitches of ~45 $\mu$m are well matched to optical detector pixel areas of 6-20 $\mu$m , and the screen sizes are larger than most detectors by factors of a few. The ability to selectively control different pixels at various illumination levels while having the rest of the screen area be completely dark is critical for testing detectors at photon-counting sensitivity levels. Perhaps most usefully, changes to the scene can be made through a programming interface, rather than by optomechanical changes. This makes it trivial to quickly switch between different testing frameworks. Temporally varying scenes are also an option to investigate effects like motion blur or transient detector responses.

---

[†]Data adapted from `www.osram-oled.com`

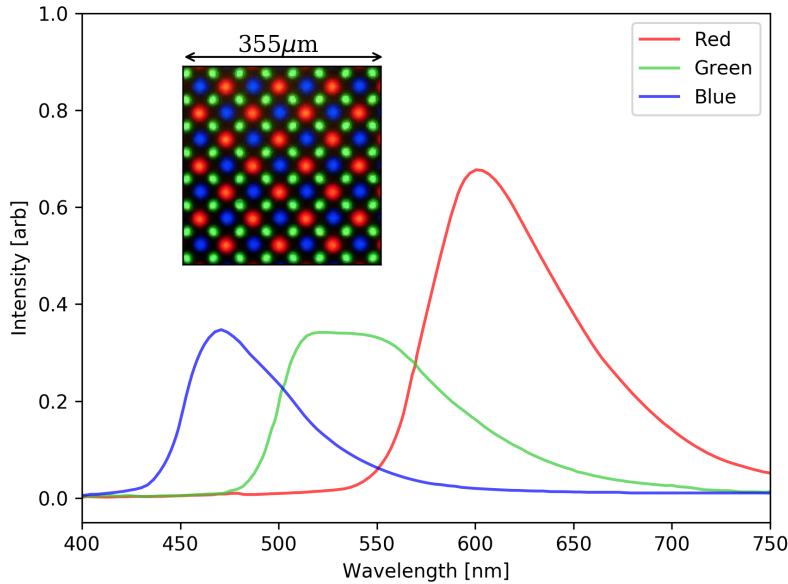

Figure 2: Representative spectra of OLED pixels. The inset shows the "Diamond Pentile" pixel configuration commonly used on Samsung smartphones.[†]

## 2.2 Optical considerations

The optical design is a two-lens reimaging system with a filter wheel and pupil stop for spot size control. The only moving part is the filter wheel, which may be omitted depending on the application. The optics were designed with the goal of being able to fill the entire detector while preserving high image quality over the field using off-the-shelf components. The Galaxy S8, the phone used in this device, has an RGB-type pixel configuration to project different colors, with 224.4 pixels/cm, the highest pixel density available when the system was designed. However, note that when manufacturers refer to pixels, they refer to the RGB unit cell, not the individual LEDs making up the unit cell. While the individual LEDs are smaller than a pixel (at about 12-22 um), the intra-LED separation for an individual color is similar to the unit cell size. For simplicity, we restricted ourselves to designing around a single LED color, green, which has the smallest intra-LED spacing.

The EMCCD has 13 micron pixels, and with a 1024x1024 active image area, this implies the maximum radial position is about 9.5 mm. The smartphone, on the other hand, has a 2960x1440 active area corresponding to a maximum radial extent of about 32 mm. To keep the tolerances and optics simple, we chose to reimage a screen area of radial extent 25.4 mm to the EMCCD active area.

For the phone collimating lens, we chose a focal length of 300 mm. In order to exactly demagnify the field onto the image area, we then would need a lens of 300mm·9.5/25.4 =112 mm. The closest compatible off-the-shelf lenses near this focal length were 100 mm and 150 mm, where we chose the latter.

A final consideration is the PSF size. At least two EMCCD pixels per spot FWHM are generally needed to properly sample the PSF. From simple geometry the expected size of the green pixel would be 12 microns multiplied by the magnification, which would be 6 microns, smaller than a single EMCCD pixel and thus undersampled. In this case, a simple solution is to introduce an pupil stop which would simultaneously control aberrations and enlarge the spot due to diffraction. Using the 545 nm central wavelength of the green LED, the spot size would be (150 mm)*(545nm)/(pupil stop diameter), which gives a 3 pixel sampling for a 3 mm iris size, easily achievable with an off-the-shelf iris. Simulations of the system in Zemax showed that this design would give diffraction-limited spots to the edge of the field (see Figure 3), with an alignment tolerance of over 1 mm.

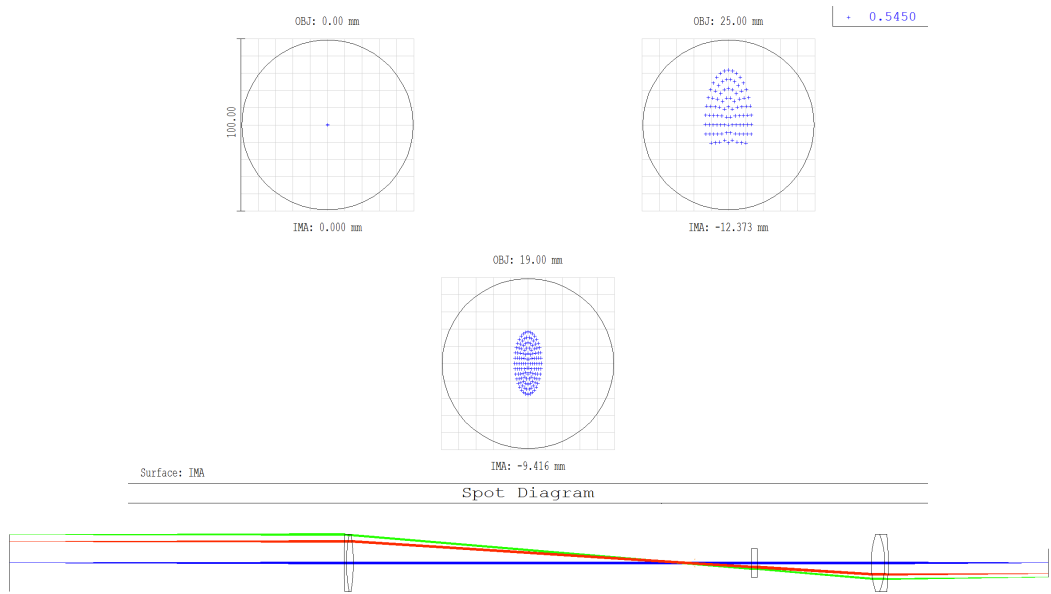

Figure 3: Optical ray trace and spot diagram of the scene generator. In the ray trace, the phone is on the left and the detector is on the right. The flat blank indicates the position of a filter wheel. The left spot diagram indicates the center of the field, the lower indicates the edge of the field, and the right indicates 3 mm past the edge of the field.

## 2.3 Mechanical design

The mechanical mounts were designed to be in a single line for simplicity. The entire system is closed, without the possibility of significant air currents or dust entering. The only custom part in the entire system is the mount for the phone, which was 3d-printed out of acrylonitrile butadiene styrene (ABS) using an industrial printer. While the 3d printing performance was not precise enough to create a truly light-tight seal, it was satisfactory and easily made light tight with some optical black tape. The base plate has extruded regions to hold the phone in place. A few features worth noting are inset o-rings to even the pressure and create a tight seal, and a hole in the side near the phone's on/off button to allow for power cycling without having to take apart the mount. While we intended to reprint or machine the mount out of metal after evaluating the design, the stability of the plastic was found to be satisfactory for our needs and so we continued to use it. The design of the mount is shown in Figure 4.

## 3. SOFTWARE DESIGN

The software we developed for this testbed can be broken down into two parts, the Android server and the Python client. The Android server is the lower-level code that controls the pixel hardware on the phone, and is written in Java. The Python client acts as the user interface for the control over the Android phone screen. At a basic user level, the phone is loaded as a module into a Python script or session, and RGB arrays of [width x height x 3] may be passed, where the width and height are the maximum resolution of the connecting phone's screen. A block diagram of the software is shown in Figure 5.

For less demanding applications, it would have been possible to get away with some available "computer-to-phone" USB link software, that essentially allows one to use mouse clicks to interact with the phone screen as one would using a finger. However, we required the low-level software control of the phone for numerous reasons. This includes the need to turn off stray lights, such as the button lights and external LEDs, that would corrupt the measurements; the ability to prevent any sort of error messages or "updates" from popping up on the screen, which are bright enough to potentially damage the detector under certain conditions; the ability to precisely control the amount of time an image is shown on the phone; and the need for continuous operation for ~week long periods.

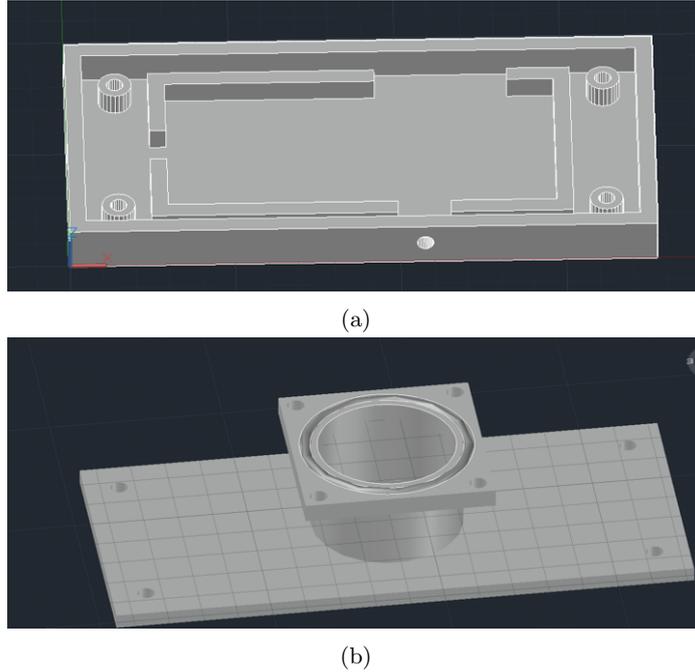

(a)

(b)

Figure 4: The two parts of the 3d printed phone mount. The cable for the phone passes through a the gap in the extruded region and hole in the mount on the left of (a), not shown.

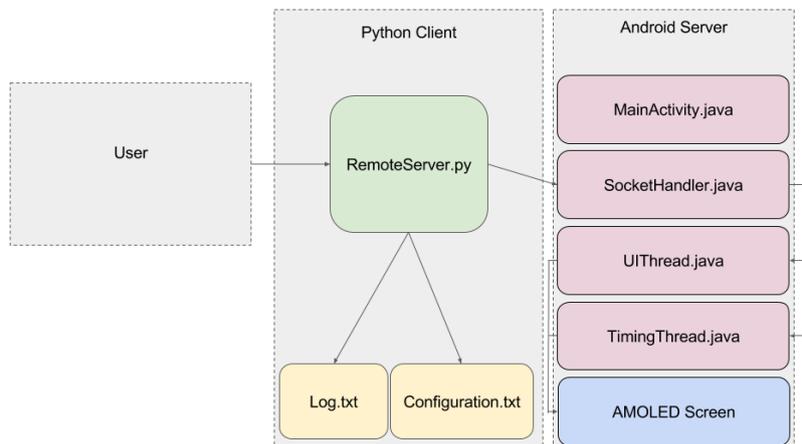

Figure 5: Block diagram of the software architecture used to control the scene generator.

We will briefly summarize the different modules of the software and how they interact with each other below.

## 3.1 Android server

In order to address the requirements mentioned above, an Android server application was developed with four main classes: the main activity, socket handler, UI thread, and timing thread. All the classes work together in a threaded control structure in order to allow the user to control the phone screen from the Python client. The Android server is written such that if an error were to arise, the phone locks the screen until the user disconnects from the Android server. This is done in order to protect the detector from getting a bright input in the form of an Android error message.

### 3.1.1 Main activity class

The main activity class contains all of the system control methods, e.g. the sequence of commands that runs when the application starts up, pauses, and shuts down. This class is normally very active in a traditional application, but our Android server did not need a control structure as it has only one state, which is a display screen. In essence, all the main activity class does is set the phone to a full screen mode, turns off notifications, and kicks off the other three classes, which are all separate threads. After the application is created and running the main activity is not used for the rest of the applications lifecycle.

### 3.1.2 Socket Handler

The Socket Handler thread is used as the communication line between the Android server and Python client. The Socket Handler begins by opening up a TCP server socket that then waits for the Python client to link to it. Once a connection is established, the Socket Handler waits for the Python client to send operational codes that correspond to different protocols, including:

- set any new screen pixel array
- set blank screen
- set screen brightness value
- set screen flash time
- disconnect from host

### 3.1.3 UI thread

The UI thread is what controls the screen's pixel values. It simply waits for an input array from the Socket handler thread. Once the array is received, it checks for correct dimensionality and values and then sends it to the screen. The thread then waits for another matrix to be received.

### 3.1.4 Timing thread

The timing thread is similar to the UI Thread as it only waits for timing data to be received. Once the timing data is received, the phone is screen is turned on and off for the specified time values that were received. This allows for accurate control of when the phone screen displays and when it does not.

## 3.2 Python client

The client is a Python "remote screen" class the user interacts with to control the scene generator. When the remote screen class is created, it looks for the signal from the Android client. The connection is made using the Android Debug Bridge's port forwarding command. This allows an Android socket port to act as a computer socket's port as long as the Android is connected to the computer via a USB cable.

Once the connection is established, the remote screen creates a byte array the size of the connected Android's screen in pixels. The pixels can then be modified by the user and then updated to the phone. This side of the software is less complex than the Android side and only has the job of sending specially formatted data, based on a custom TCP protocol, that the user requests.

The Python client also stores both a log and configuration files that record data for debugging and data analysis. The configuration file saves all of the configuration data and the state of the current screen at the beginning of a connection. The log files contain all of the commands that were used during the most recent streaming session. This allows the user to recover the activity during a testing period.

# 4. PERFORMANCE

Before interfacing the device to our science-grade EMCCD dewar, we commenced testing with an Andor iXon Ultra camera, which uses the same EMCCD chip but different readout electronics. Despite the common detector chip, there were some notable differences in our tests. The initial testing with the iXon was not done in a controlled environment, but on an office table and not mounted to any sort of optical breadboard. As such, there was noticeable image motion due to temperature swings, people walking by, and other annoyances.

## 4.1 Dark levels

A first and critical test was to make sure the phone dark levels were consistent with zero. This test is important to guarantee that there is truly no extra light generated by the phone.

We performed two separate measurements. On the iXon test camera, we powered the phone off and integrated for 2000 seconds. Second, we powered the phone on and sent a screen consisting of all zeros, which corresponds to having the OLEDs turn completely off, then took another 2000 second exposure. We compared image histograms and found no statistically significant difference.

Later, we ran a similar test with the science-grade EMCCD, where we ran hour-long exposures at a very high sensitivity mode and compared images. This again revealed no stray light in the system. The most stringent tests were "performed" when we later ran experiments lasting for days in photon counting mode. In this scenario, the camera gain is very high and the expected photon rate from the target pixel is well less than one photon/frame. In this sort of test, any stray light from other pixels of the phone will easily overwhelm the tiny signals we are trying to measure. Again, no stray light was detected. These tests confirmed that pixels on the phone screen do not produce any light when turned off, consistent with the expectations of the OLED technology.

## 4.2 Illumination stability and uniformity

Another important test was to determine the level of illumination stability and uniformity in the device. Our intended application required running tests for days to a week without changes in scene brightness. The stability can be characterized by the image motion, photometric precision and accuracy, and PSF uniformity. There was no guarantee that the screen would be able to perform at a high level, as it was built for human eyes which cannot distinguish subtle changes in these parameters during normal phone use.

### 4.2.1 Pixel motion

Limiting image motion is important for our experimental goals. As our experiment flux levels are below 1 photon per frame, it is impossible to determine image motion from the experimental data itself. We determined that drifts of 1/10 of our FWHM (that is, less than 1/4 pixel) could be tolerated. We note that the optomechanical setup, not the phone screen, controls the image motion.

To test image motion, we generated a grid of bright spots (similar to the inset in Figure 7) and tracked their motion over multiple hours. The pixel motion stability exceeded our needs. Over many hours, the drift was well below 1 $\mu$m (0.07 pixels), with periodic spikes of ~0.4 $\mu$m every ~25 minutes from the automatic nitrogen top off. The system transient settling time response after being started is about 1 hour (see Figure 6) Over periods of many months, we measured a drift of approximately 10 microns. This may be due to the first author of this work constantly bumping the device, settling, or temperature swings. The lab is not climate controlled to better than a degree, so some thermal breathing is to be expected.

### 4.2.2 PSF shape

While our particular application does not require high PSF uniformity, this can be important for some applications. We measured the PSF shape by generating a grid of ~200 points over the entire field of the detector, and fit ellipses to each PSF. We found that the ratios of semi-major to semi-minor axes of the ellipses to be 1.01±0.02, indicating a 1±2% deviation from perfect circularity. This is likely an upper bound, as it is likely that this is dominated by the SNR of the measurement or detector readout "smearing" effects. Regardless, this small level of asymmetry is negligible for our interests.

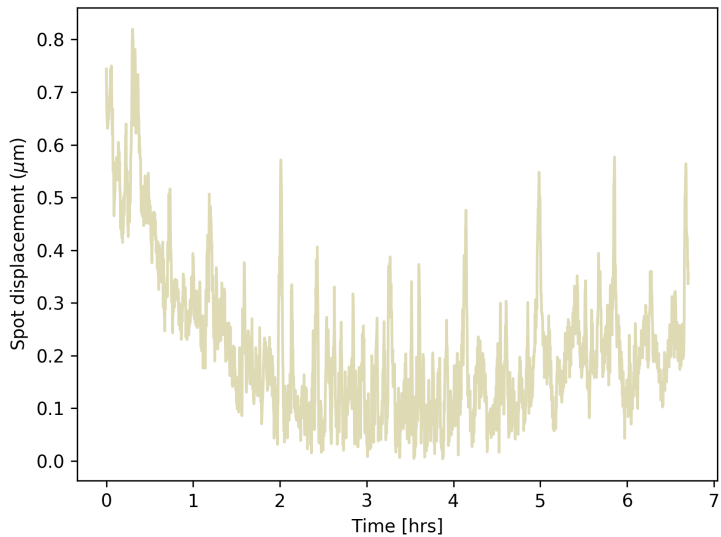

Figure 6: Image motion over a ∼7 hour period immediately following a cooldown. The plotted line is the median displacement from the nominal position of a grid of points generated on the phone and imaged by the detector. The system transient response dissipates after approximately one hour, with 0.4 $\mu$m spikes due to the cryocooler fills visible every 25 minutes or so.

### 4.2.3 Photometric stability and uniformity

A critical factor is the photometric variability of the pixels. If pixels drift significantly in intensity, our experiments would not be possible. To test this, we generated a grid of spots of various intensities (color values from 0 to 255) and tracked their brightness (using aperture photometry); taking 20 second exposures continuously for 3 days. We were pleasantly surprised to find that the spot brightnesses were extremely stable, nearly to the Poisson limit (see Figure 7). There was one exception to this; when setting pixels on the phone to the brightest possible settings (RGB, brightness values of >220), we noted some persistence effects in brightness that would last for some hours after the pixels were returned to lower brightness values. This was not found at intermediate and faint settings.

The interpixel uniformity, however, was poor. Different pixels set to the same RGB value were typically only photometrically equivalent to $\Delta I/I$ of 20%, with excursions of 50% not uncommon at high RGB brightness. As the human eye reacts to intensity logarithmically and cannot even resolve the separate LEDs in a single phone pixel, it is unsurprising that the phone screen is not calibrated to a high degree of pixel uniformity. However, this does leave open the option of a user-defined calibration bringing down the interpixel variance to better than 10%.

### 4.3 Refresh rate and dynamic range

The high flexibility of the smartphone screen opens the possibility of testing detector performance on moving scenes. We did not investigate this potential in great detail as most of our experimental interests are in static, differential measurements. However, we do occasionally use "pre-flash" effects, which involves setting a uniform illumination on every pixel (a flatfield), exposing and clearing, and then immediately resetting the science scene for a longer exposure with the detector. Such experiments require knowledge and control of the transition speed between the different exposures.

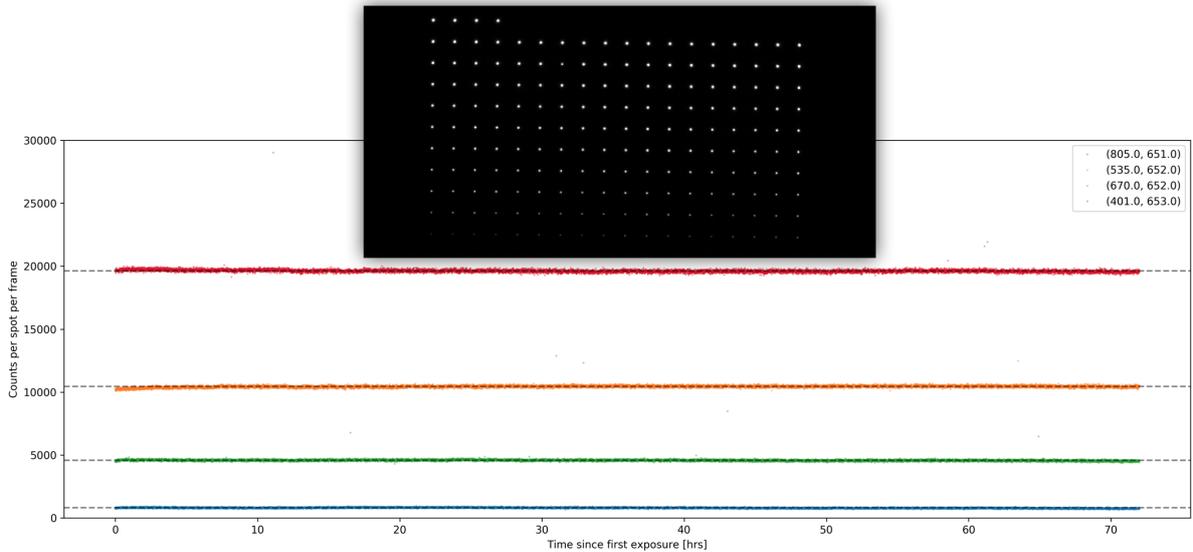

Figure 7: Photometric stability over a 30 hour period. A grid of points at RGB = (0, 0-255, 0) were generated (upper inset) and aperture photometry was performed on all the points. Time series of four points are shown. This data was taken with the test camera; as such, image stability was not very high and the effects of settling on the photometry are visible in the first few hours. The occasional spurious $> 10\sigma$ spike is caused by the aperture photometry algorithm interpreting cosmic ray events as the PSF, and is not a real brightness variation.

#### 4.3.1 Refresh rate

The limit to the refresh rate of almost all smartphone screens is 60 Hz, so this presents an upper limit to what could be achieved without low-level hacking of the I/O. In addition to hardware limitations, there are software limitations. We load large bitmap images on the phone screen, so data transfer speeds and programming optimizations are relevant. With our software it takes about 0.4 seconds for a full image to load. This can be improved by compressing the image to a jpeg rather than a bitmap, at which point it only takes a few hundredths of a second for the image to load. However, with a compressed file, the photometric accuracy and image shape can degrade. This was a trade we were not willing to make; though for other applications this may be acceptable.

#### 4.3.2 Dynamic range

The dynamic range of the phone screen refers to the maximum flux difference between two different pixels during a single exposure of a single screen. To determine the dynamic range, we set the phone pixels to a range of 0-255 and took ten second frames for 3 days (this is the same dataset as shown in Figure 7, using the iXon test camera). We performed minor preprocessing to remove systematics like bias and took the temporal average of the dataset. We then performed aperture photometry on a representative sample of the faintest and brightest few spots that were visible in the image (only 207 spots were visible). Aperture photometry of the faintest spot returned 1.39±1.36 counts, but it is possible that this spot corresponded to a bad (phone screen) pixel, so we neglect it. The next faintest spots were measured to contain 6.12±2.48 and 6.45±2.53 counts. The brightest spot returned 31,068±177 counts. These numbers indicate a dynamic range of 6400±4270 or about 37±2 dB. It is possible that there were even fainter spots below the noise that we could not see, in which case the dynamic range could be even higher.

### 5. DISCUSSION AND CONCLUSION

We have presented the design and performance of a scene generator for characterizing visible imaging detectors. Using a modern smartphone, the instrument is a flexible device that is appropriate for a range of applications, including the demanding characterization of space-based exoplanet imaging detectors. The scene generator can be constructed at modest cost and only requires one custom part; a mount for the phone.

We provide a few examples of what the scene generator can do in the Section 6.2, showing demonstrations of a flatfield, a uniform diagonal grid, and a complex scene simulating the WFIRST integral field spectrograph.[5] This shows the capabilities for generating standard test patterns as well as complex astrophysical scenes with high fidelity to expected data.

Over the ~year of use of this system, reliability and reproducibility have been excellent. The one problem we have encountered was the battery beginning to lose its ability to hold charge. This is due to using it continuously for months, and showed up as the phone powering down during times of heavy screen use. The solution we are investigating is getting a more powerful charging USB power hub. Alternatively, we will just replace the phone with a newer one.

# 6. APPENDIX

## 6.1 Parts list

| Item | Part No. | Qty | Total cost (USD) |
|---|---|---|---|
| SM2 Lens Tube Without External Threads | SM2M30 | 3 | 100 |
| SM1 Coupler, External Threads | SM1T2 | 2 | 40 |
| SM2 Lens Tube, 1" Thread Depth | SM2L10 | 2 | 60 |
| Adapter, External SM2->Internal SM1 | SM2A6 | 1 | 25 |
| Ring-Actuated SM2 Iris Diaphragm | SM2D25D | 1 | 85 |
| Adapter, External SM1->Internal C-Mount | SM1A10 | 1 | 20 |
| SM2 Lens Tube, 3" | SM2L30 | 7 | 260 |
| SM2 Lens Tube, 0.3" | SM2L03 | 1 | 24 |
| 2" Achromatic Doublet, f= 150 mm | AC508-150-A | 1 | 106 |
| 2" Achromatic Doublet, f = 300 mm | AC508-300-A | 1 | 106 |
| SM2 Coupler | SM2T2 | 2 | 70 |
| Samsung Galaxy S8 | N/A | 1 | 630 |
| 3d-printed phone mount | N/A | 1 | 250 |
| Filter wheel (optional) | FW102C | 1 | 1100 |
| **Totals** | | | 1776 |
| Totals (with filter wheel) | | | 2876 |

Table 1: All part numbers from Thorlabs. 3d-printed mount made on-site at the JPL machine shop. Parts for mounting on optical breadboard not included

## 6.2 Application examples

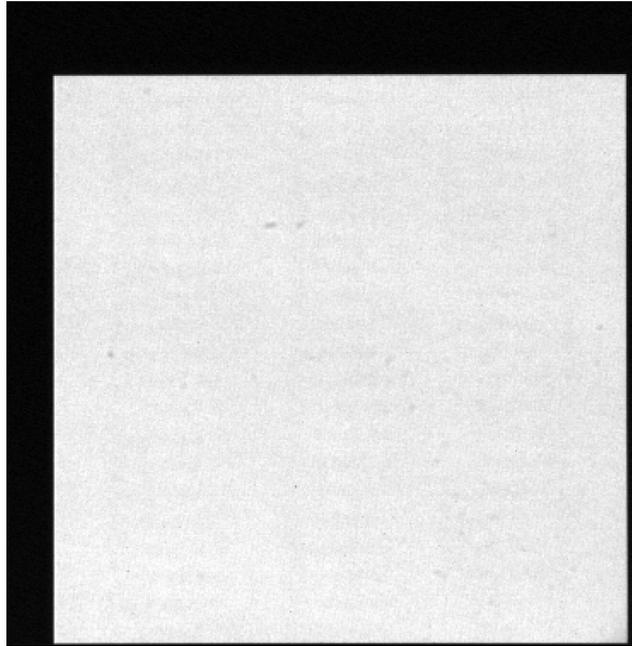
Figure 8: A flatfield image is generated over the active detector area. This flatfield image is of a single channel, which delivers a 3% flatfield accuracy (the individual pixels of the phone are not perfectly resolved and cause a subtle sinusoidal oscillation in the flat). A better flat can be had by combining and balancing all three channels, as the three RGB color channels do not overlap spatially, or including a diffuser in the filter wheel. A portion of the CCD overscan is visible in this image.

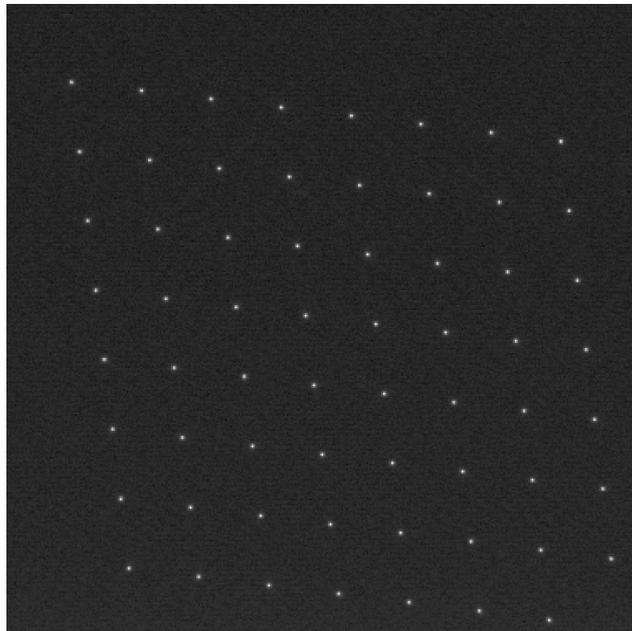
Figure 9: Scene generator creating spot grid of PSFs. The diagonal offsets guarantee that each PSF occupies its own row and column, allowing for effects from the vertical and horizontal transfers of the CCD readout electronics to be independently investigated. A portion of the CCD overscan is visible in this image.

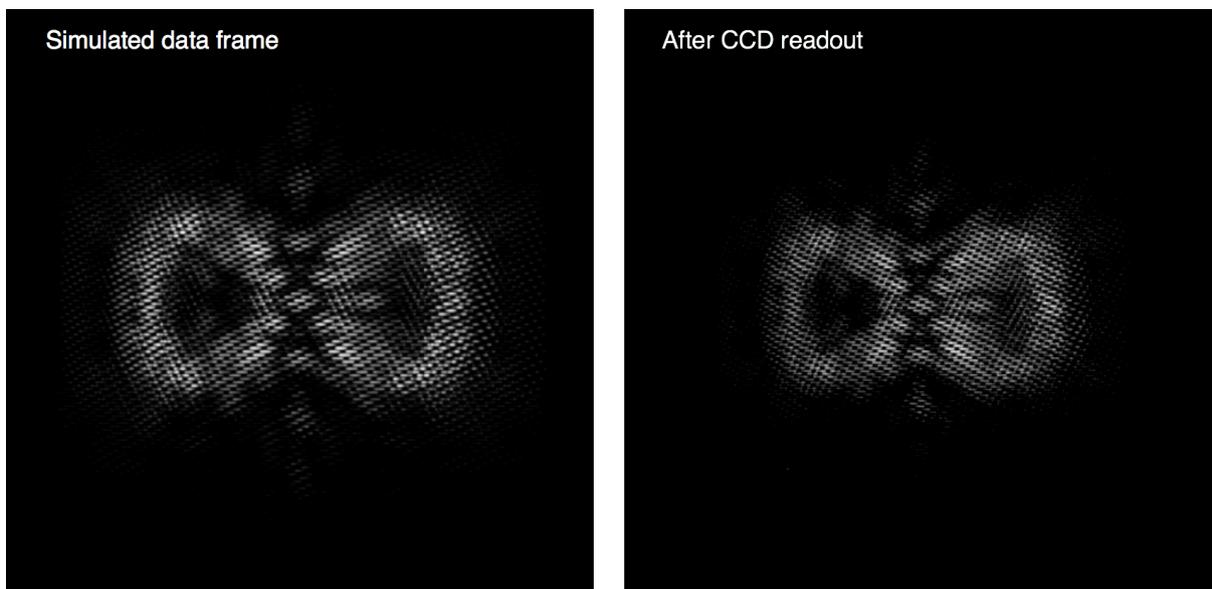

Figure 10: The image on the left is a simulation of the data expected from the WFIRST integral field spectrograph.[5] The image on the right is an image obtained after loading the simulated image into the scene generator, performing a spatial and intensity calibration, projecting it onto the EMCCD, and reading out the EMCCD. Note the calibration is not perfect, so there are discrepancies between the size and intensity of the images. However, the calibration can be improved in software to better reproduce the expected data.

# ACKNOWLEDGMENTS

We thank J. Kent Wallace and Eric Cady for advice and material support towards this work, and to Maxime Rizzo for WFIRST integral field spectrograph data. We thank Ben Mazin for first making us aware of the potential of OLEDs for detector characterization. This work was performed at the Jet Propulsion Laboratory, California Institute of Technology, under a contract with the National Aeronautics and Space Administration, (C) 2018. All rights reserved. Government sponsorship acknowledged.